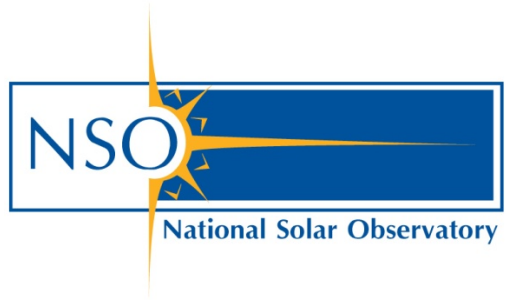

# Image Quality of SOLIS/VSM in Helium vs. Nitrogen

J. W. Harvey

National Solar Observatory

May 22, 2014

### Abstract

The National Solar Observatory (NSO) Synoptic Optical Long-term Investigations of the Sun (SOLIS) Vector SpectroMagnetograph (VSM) is sealed and was designed to be filled with helium at slightly above ambient pressure. After 11 years of operation filled with helium, an acute shortage of helium prompted a test using nitrogen as the fill gas. Four months of nitrogen-filled observations in 2014 are compared the same months in 2013 with helium fill. On average, the image sharpness is slightly degraded when using nitrogen.

**Introduction.** The SOLIS VSM is a 50 cm aperture telescope equipped with a long-slit spectrograph that scans a full-disk solar image with spatial sampling of 1 arc sec and a variety of spectral sampling values (Keller et al. 2003). Its principal products are measurements of magnetic fields on the solar surface and in the solar chromosphere. The instrument is sealed, and a 74 cm diameter, 6 mm thick fused silica window allows entry of sunlight. The telescope is a quasi-Ritchey-Chretien design with a primary mirror operating at f/1.6. The ~ 400 W of solar light from the primary is reflected by a secondary mirror fabricated from a single silicon crystal. The final f/6.6 full-disk solar image is focused on a spectrograph slit that is cooled by a flow of chilled water-propylene glycol solution. The mirrors are coated with protected silver.

To reduce internal seeing and to cool heated optics, the instrument was designed to be filled with helium gas at near ambient temperature and pressure. This also protects the mirror coatings, which otherwise degrade rapidly when exposed to water vapor and oxygen. Tests during the design phase showed that the forced convective flow of helium cools a surface about as efficiently as dry air. The most desirable property of helium is its low index of refraction (13% that of air), which means that thermal fluctuations along the optical path are much less disturbing to image quality than when using other gasses. Furthermore, helium has about five times the specific heat capacity of air. Other solar telescopes filled with helium include the IVM (Mickey et al. 1996) and THEMIS (Arnaud et al. 1998). The virtues of helium-filled solar telescopes have been discussed in detail elsewhere (Engvold and Andersen 1990; Engvold et al. 1981, 1983; Mehltretter 1981; Nielsen 1992; Rösch 1965).

It is well known that helium gas is becoming scarce and expensive (NRC 2010). At the end of 2013, NSO was notified by its supplier that they were unable to provide helium as readily as during the previous 11 years. This prompted consideration of using gases other than helium. Among these were hydrogen, neon, nitrogen, and various 'shielding gasses' used for welding that are mixtures of helium and argon. Several considerations led us to try dry nitrogen as a replacement for helium. But we were concerned about detrimental effects on cooling and image quality – leading to this study.

**Image sharpness from Doppler observations.** Several full-disk solar observations are made daily using the VSM. Among the data products produced are maps of the Doppler shift of the 630.15 nm Fe I and 854.2 nm Ca II lines. These are produced at level 1, which means that numerous geometric corrections have not been made. Figure 1 shows two of these maps. The vertical gap in the images is a result of splitting the image between two cameras.

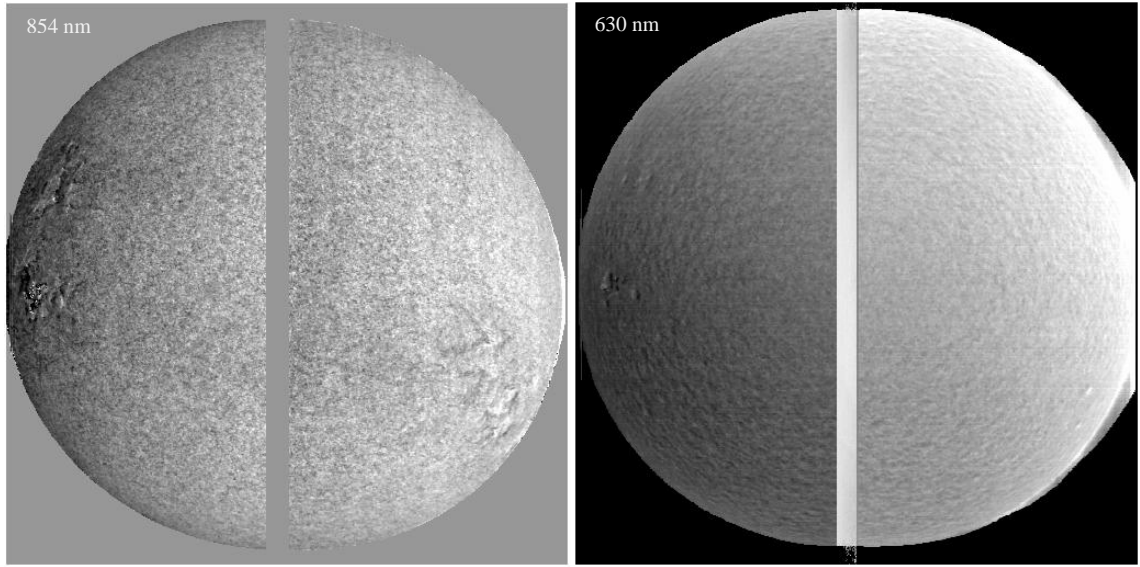

Figure 1. Level 1 VSM maps of Doppler shifts measured with 854.2 nm Ca II (left) and 630.15 nm Fe I (right). January 29, 2014.

In these images, in addition to a large gradient due to solar rotation and some instrumental offsets, there are small-scale structures present everywhere on the solar disk, though these structures have a different character in the 854 and 630 nm images. The visibility of this small-scale structure is a good candidate as a measure of image sharpness. Note that these images are built up by scanning the solar disk with a spectrograph slit. Each row in the image represents simultaneous measurements of the Doppler shifts along the spectrograph slit. The exposure times for each line are 1.28 and 0.32 seconds, respectively, for the 854 and 630 nm images. Thus, it requires 2621 and 655 seconds, respectively, to build up entire images of 2048 rows.

The Doppler images can be filtered to enhance the small scales and suppress large-scale fluctuations. This is done separately for each row since image quality can vary rapidly row-to-row due to atmospheric fluctuations (seeing). The filter adopted here is a convolution along a row with a three-element kernel that approximates a second derivative of the Doppler shift signal along the row, namely values -0.5, 1.0, -0.5. In order to deal only with positive numbers, the absolute value of the convolution is used hereafter. Better image sharpness leads to larger values. To reduce unwanted effects of a changing chord length across the image and center-to-limb variations of solar structure, the average value of the convolution in each row is computed only for several hundred pixels on either side of a column passing through the center of the solar image. A few persistently noisy columns and portions of the solar disk that are unusually dark or bright are also excluded from the averaging. The result is a single number for each image row that represents image sharpness for the time that row was being observed.

While the sharpness value cannot easily be calibrated in terms of image blurring in units of, say, arc seconds, comparisons are possible since the way it is calculated is always the same. To get a very rough calibration we start with a good 630 nm observation having an initial Gaussian one-sigma blurring estimated as 1.4 pixels (or arc seconds) and blur it additionally by convolution with Gaussians having sigmas of 1, 2, or 4 arc seconds. From these results we find that $b^2 \approx a/(s^{1/2} - c) - d$ where $b$ is the additional blurring in arc seconds, $s$ is the sharpness signal, $d$ is the square of the initial blurring (here 2), *c* is a constant (here 0.35) and *a* is another constant (here 5.3). So sharpness values of 9, 4, 2 lead to additional blurring of 0, 1.1, 1.7 arc seconds respectively. A similar non-linear relation is expected to apply to 854 nm observations.

Figure 2 shows the time variation of the sharpness signal during scans on days of good and poor seeing conditions using the two spectrum lines.

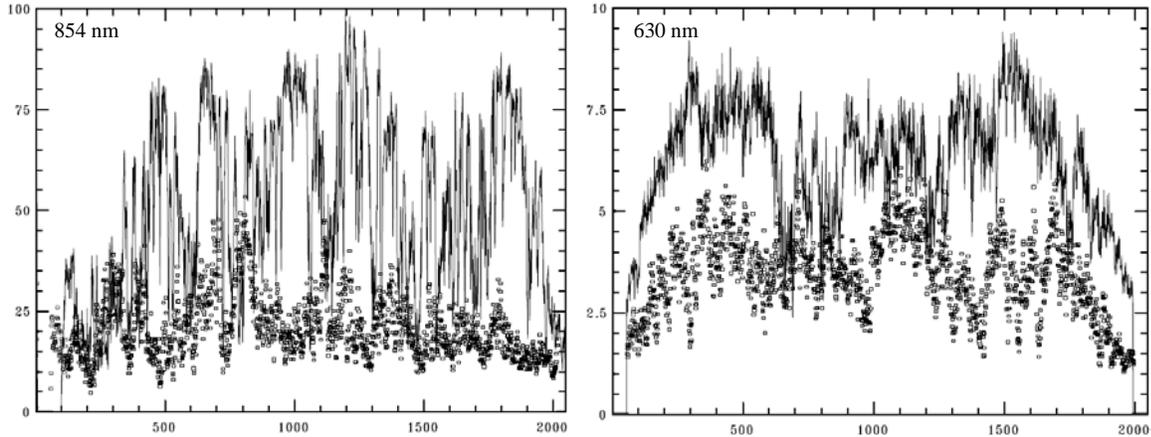

Figure 2. Sharpness signals as functions of scan line number (equivalent to time) for 854 nm observations (left) on 2014 January 20 (boxes) and 21 (lines) and 630 nm observations for the same days (right). The fall off of the signals near the edges is a geometric effect.

It is interesting to note that the statistical distribution of the sharpness values is not Gaussian but is more nearly a log-normal distribution, though with a lot of variability. Thus, statistical measures of sharpness such as mean, median, etc. may not best represent the average or typical image quality in the most appropriate way. Better might be to use values of quintiles or some similar measure of histograms as we show later. The sharpness values tend to be spiky in time with excursions between good and bad image quality over periods of a few minutes at the Kitt Peak location of the SOLIS VSM. Nearly simultaneous measurements with a scintillation monitor (Seykora 1993) mounted at the edge of the SOLIS VSM aperture did not correlate well with the actual image sharpness (L. Bertello, private communication). The reasons for this discrepancy are likely due to the limited distance range from the sensor over which scintillation produces a measureable signal (Beckers 1993), the need for very close synchronization in time of the scintillation and sharpness signals (Coulter et al. 1996), and the low cadence of the scintillation sampling used compared to the image sharpness signal.

**Stack plots.** Stacking sharpness signals like those in Figure 2 into a 2-D data plot is a convenient way to store and assess a lot of data. This was done for 630 and 854 nm full–disk observations taken between January 22 and May 19 in 2013 (helium filled) and 2014 (nitrogen filled). Observations during clouds were excluded. The 630 nm observations are more frequent than the 854 nm ones. Figures 3 and 4 show the results. The plots are shown with identical gray scales and the generally brighter signal in the top panels shows that sharper images are more frequent with the helium-filled instrument. We note that extended periods of better and poorer sharpness lasting many days are seen in the figures. These periods appear to be correlated in both 630 and 854 nm observations.

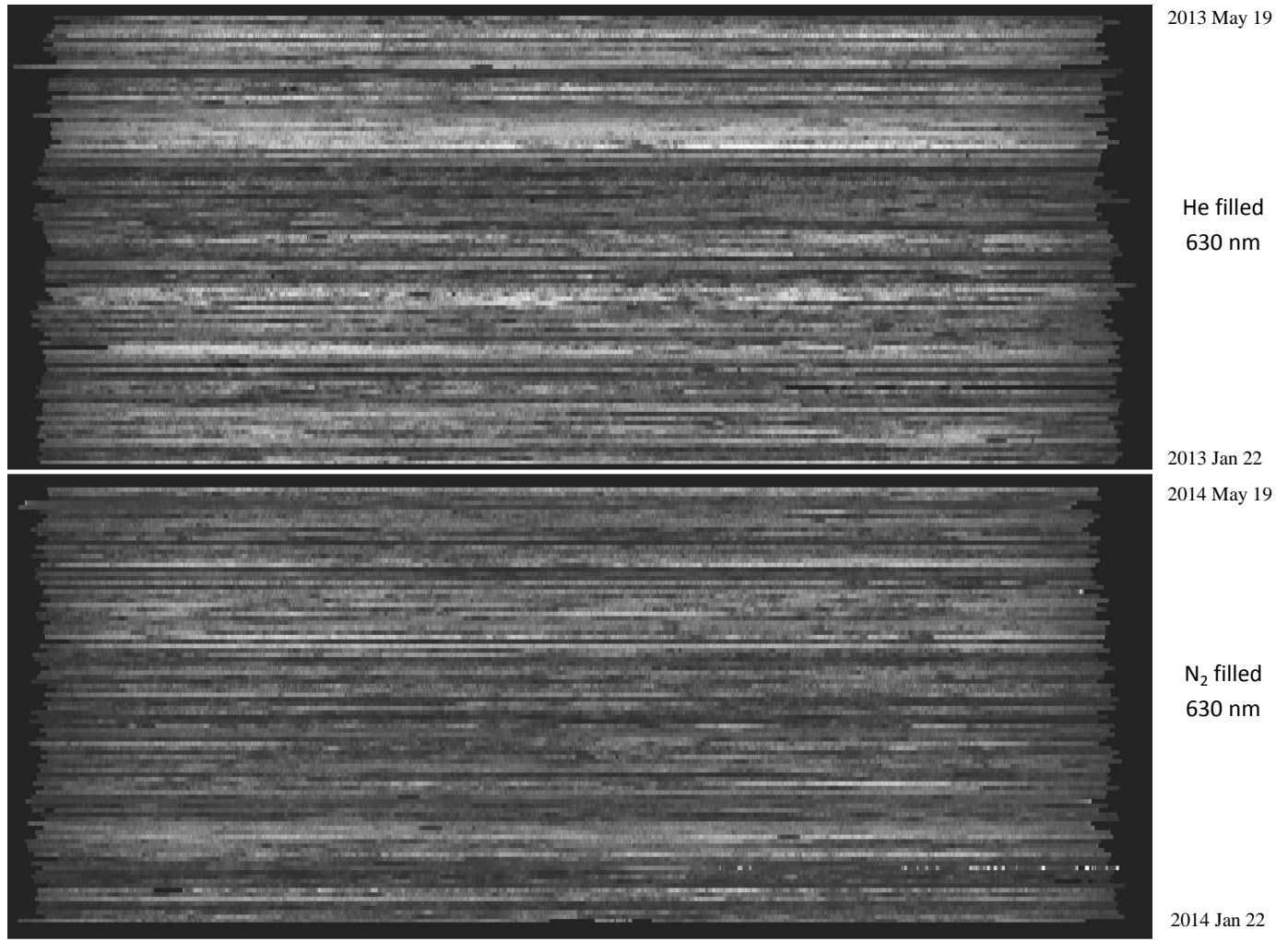

Figure 3. Stack plots of sharpness values from 630 nm observations (helium filled on top and nitrogen filled on the bottom). Abscissa is scan number (or time) increasing from left to right and covering about 10 minutes. Calendar date increases upward in each panel. Lighter indicates sharper.

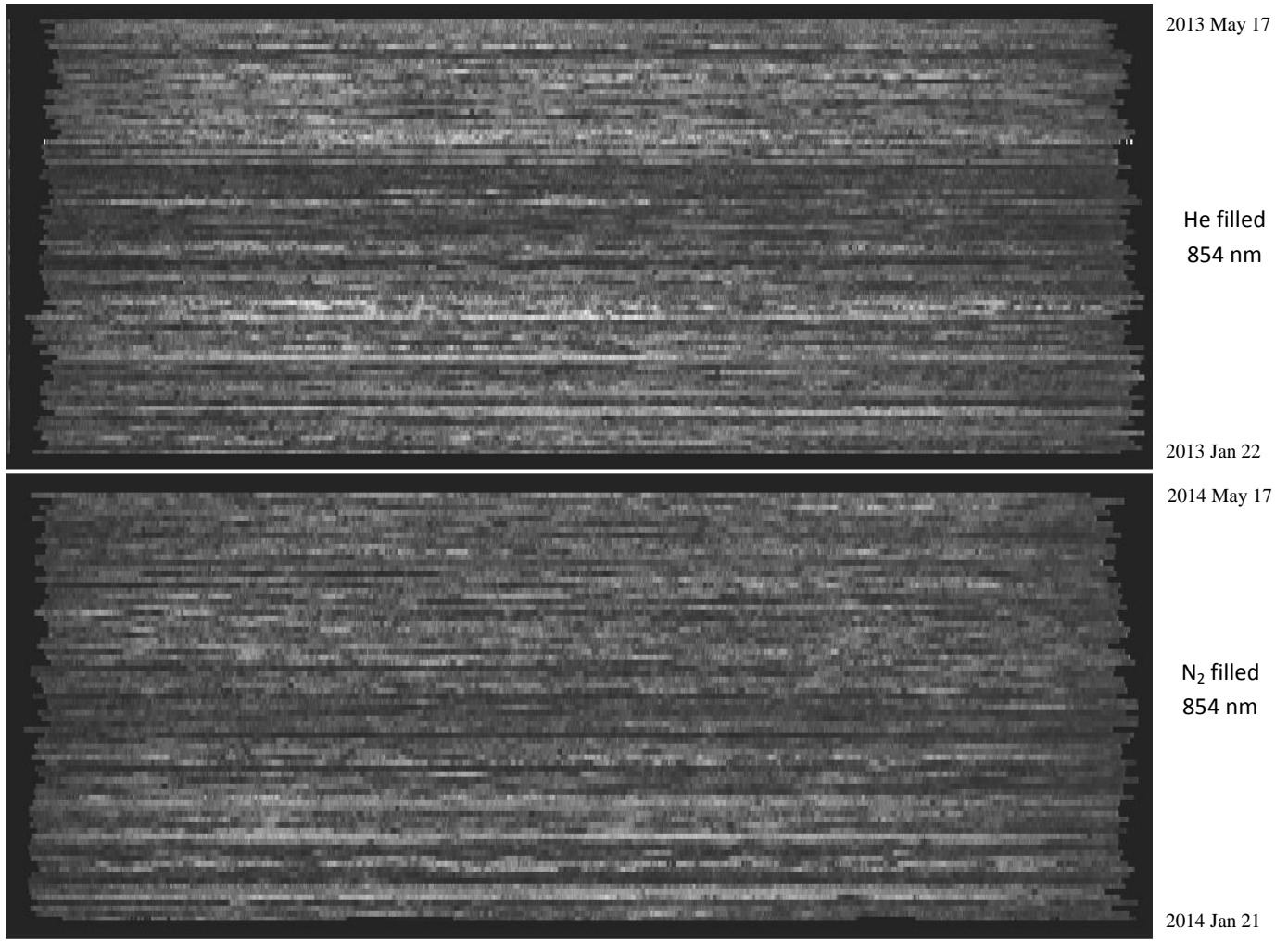

Figure 4. Stack plots of sharpness values from 854 nm observations (helium filled on top and nitrogen filled on the bottom). Abscissa is scan number (or time) increasing from left to right and covering about 40 minutes. Calendar date increases upward in each panel. Lighter indicates sharper.

As mentioned earlier, the complicated distribution function of the sharpness values of one observation makes it difficult to assign a single number for the sharpness of an observation. To deal with this problem, histograms of the data are constructed as gray-scale images in a format similar to that of Figures 3 and 4. Edge effects are avoided by excluding the first and last 300 measurements of each observation. The results are shown in Figure 5.

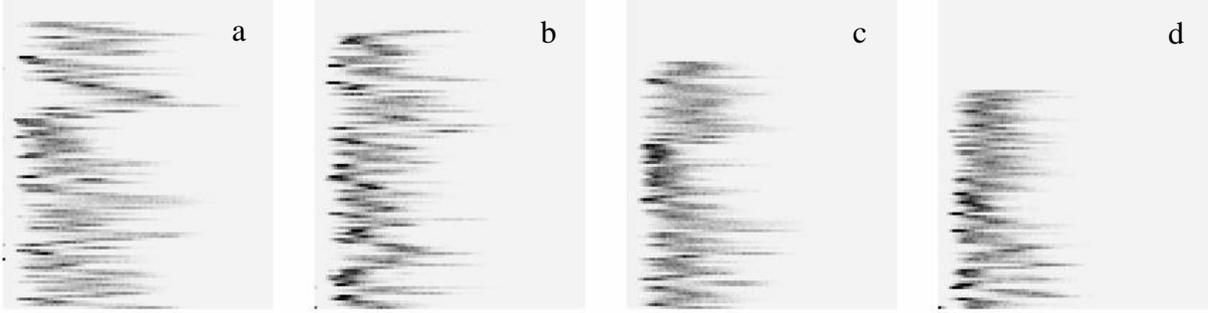

Figure 5. Histograms of sharpness values included in Figures 3 and 4 shown in the same vertical format as those figures. In each panel, the sharpness values increase to the right and darkness indicates the number of measurements in each of 100 sharpness bins. Calendar date increases upward. (a) 630 nm helium filled in 2013. (b) 630 nm nitrogen filled in 2014. (c) 854 nm helium filled in 2013. (d) 854 nm nitrogen filled in 2014.

In Figure 5, the sharpest observations are those with large histogram counts toward the right side of the panels. To the contrary, a large histogram count near the left edge corresponds to a large number of fuzzy images. It is useful to compare the averages of these histograms for the helium and nitrogen-filled time periods as shown in Figure 6.

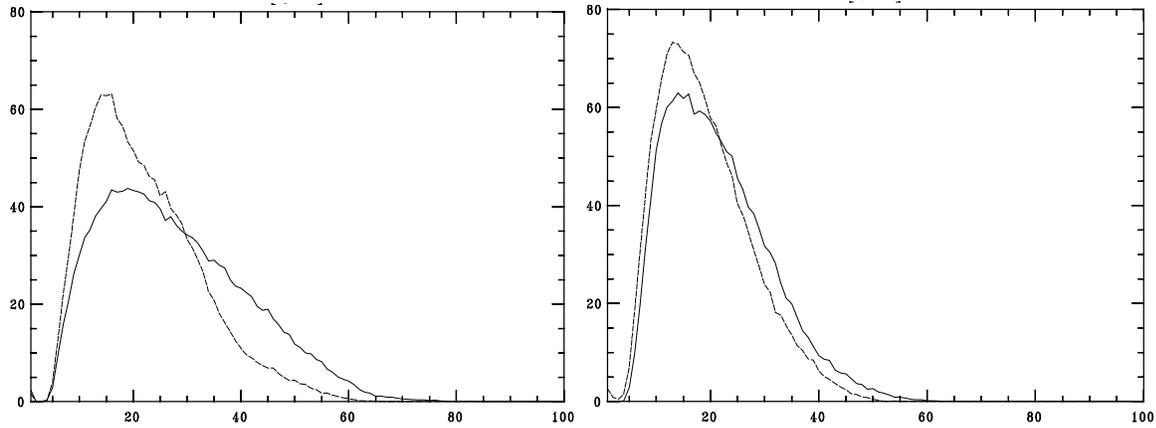

Figure 6. Average sharpness histograms for 630 (left) and 854 nm (right) observations. Solid lines are for observations made with helium gas in the instrument and dashed lines for nitrogen fill. The abscissa is bin number and the ordinate is average counts within the bin. Bin number 100 corresponds to sharpness values of 20 and 150 for 630 and 854 nm observations respectively.

The average histograms show superior average sharpness for the helium-filled instrument. Figures 3 and 4 show the same thing.

Figures 3 and 4 show what appear to be differences between the time fluctuations of the seeing in the 630 and 854 nm observations. There is a difference of a factor of four in the time sampling between these observations. We may expect that there might be differences of the time behavior of internal seeing when the instrument is filled with helium or nitrogen. To investigate this possibility, frequency power spectra were computed for each observation shown in Figures 3 and 4. The first and last 10% of each observation were apodized by a cosine bell function and the mean was not subtracted in order to leave low frequencies minimally affected. The results are shown in Figure 7.

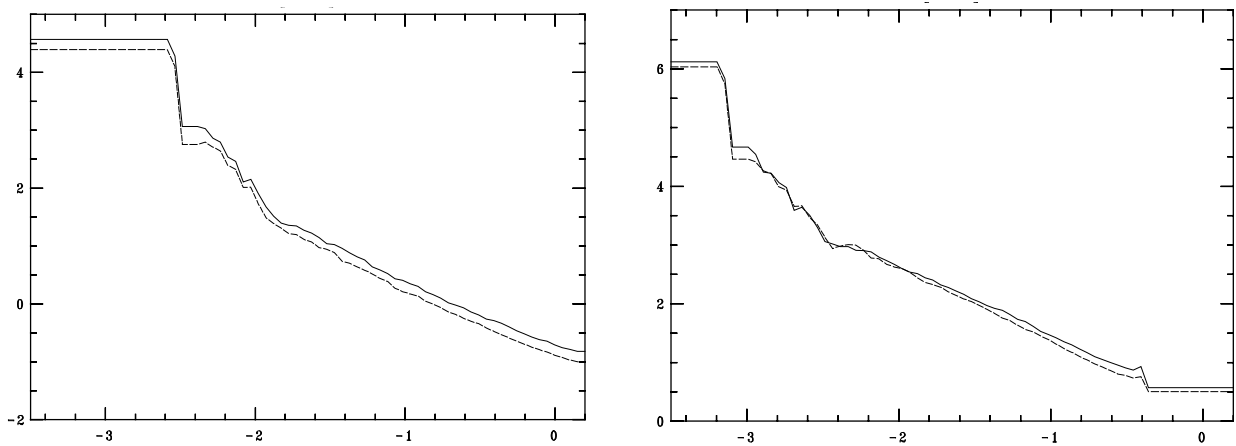

Figure 7. Average power spectra for 630 nm (left) and 854 nm (right). Abscissa is $\log_{10}$ frequency in Hz and ordinate is $\log_{10}$ power. The steps in log frequency are 0.05. Solid lines are for helium-filled-instrument observations and dashed for nitrogen-filled.

The data in Figure 7 show that the helium-filled-instrument observations are sharper than when the instrument is filled with nitrogen, as seen in previous results in this report. An exception is the near equality for 854 sharpness power in helium and nitrogen atmospheres between frequencies of 0.01 and 0.001 Hz. The changes in slope at -1.9 (for 630 nm) and -2.4 (for 854 nm) are likely artifacts associated with the data apodization since the slope changes occur at different frequencies but at about the same fractions of the sampling frequency. The average value of about -1 of the slope at frequencies above 0.01 Hz is surprising since power spectra of meterological quantities typically vary with frequency as -5/3. Correcting the aliasing of power from frequencies above the Nyquist frequency would steepen the slope a small amount at high frequencies..

**Discussion.** Based on four months of observations with the SOLIS VSM filled with nitrogen rather than helium we conclude that there is a clear, systematic, small degradation of image sharpness. However, the change is smaller than day-to-day variations in sharpness when filled with either gas. An important limitation that is hard to isolate is the quality of the telescope focus done by the observers prior to each observation. The data analyzed here were collected during a generally unfavorable time of year for obtaining sharp images. It may be that better external observing conditions would show an instrumental upper sharpness limit that is less with nitrogen than with helium. Using nitrogen instead of helium is certainly cheaper but does reduce image quality. This study did not look at differences in the cooling of internal telescope components effected by forced flow of the two different gases.

**Conclusion.** If availablity and cost are not considerations, helium should be used as a fill gas in order to obtain better image quality. The improvement is more significant for observations near 630 nm than for 854 nm. While quantitatively small increases of image sharpness may seem to be of small value, it is worth pointing out that a measured magnetic flux density for an isolated unresolved feature varies as the square of the image sharpness.